\documentclass[aps,prl,twocolumn,showpacs,showkeys]{revtex4-1}

\usepackage{graphicx,color,amsmath,bm,amssymb,dcolumn}
\usepackage[USenglish]{babel}

\begin{document}

\title{The absence of surface states for LiFeAs}

\author{Alexander Lankau $\dagger$, Klaus Koepernik, Sergey Borisenko,
  Volodymyr Zabolotnyy, Bernd B{\"u}chner, Jeroen van den Brink and
  Helmut Eschrig}

\email{h.eschrig@ifw-dresden.de}
\homepage{http://www.ifw-dresden.de/~/helmut}

\affiliation{IFW Dresden, PO Box 270116, D-01171 Dresden, Germany}

\begin{abstract}
We investigate %
the cleaving behavior of LiFeAs and determine its surface %
electronic structure by detailed density functional calculations. %
We show that %
due to the neutral surface of LiFeAs after cleaving, barely any %
influence of the surface on the electronic states is present. Therefore the data of surface sensitive probes %
such as angle resolved photoemission spectroscopy (ARPES) %
represent to a high degree the bulk electronic structure. This we highlighted by a direct comparison of the calculations to ARPES spectra. 
\end{abstract}


\maketitle

The use of angle-resolved photoemission (ARPES) and scanning
tunneling spectroscopies to study the electronic structure of
iron pnictides can turn out to be crucial to understand the
superconductivity in these materials. On one hand, both methods are very
informative by providing the detailed distribution of electron density in real
and reciprocal spaces, but on the other hand it is not clear whether the
obtained information is related to the bulk or surface.

In a recent paper \cite{eschrig2} we found a pronounced surface electronic %
structure in LaOFeAs, which appears to be present in published angle-resolved %
photoemission (ARPES) data.\cite{Lu08,ARPES1,ARPES2,AMES,YangArxiv,Liu10} LaOFeAs single crystals cleave between the %
La and As layers forming a polar surface. In contrast LiFeAs, of which clean %
large single-crystals are now available \cite{haus,song,tappl}, can be expected to cleave %
between adjacent Li layers and hence to produce a neutral surface. %
This is confirmed by results of density functional (DFT) calculations %
presented below.%

Although LiFeAs in bulk is found much more three dimensional (3D) compared %
to LaOFeAs in DFT calculations \cite{eschrig1,miyake10,nakamura09}, its surface electronic %
structure is much more bulk-like. Surface sensitive probes %
are therefore expected to directly yield information on the bulk behavior. %
Also, undoped stoichiometric LiFeAs is superconducting below 18~K and %
does not order magnetically.\cite{tappl,chua}

\begin{figure}%[ht]
  \centering
  \includegraphics[scale=0.5,clip=]{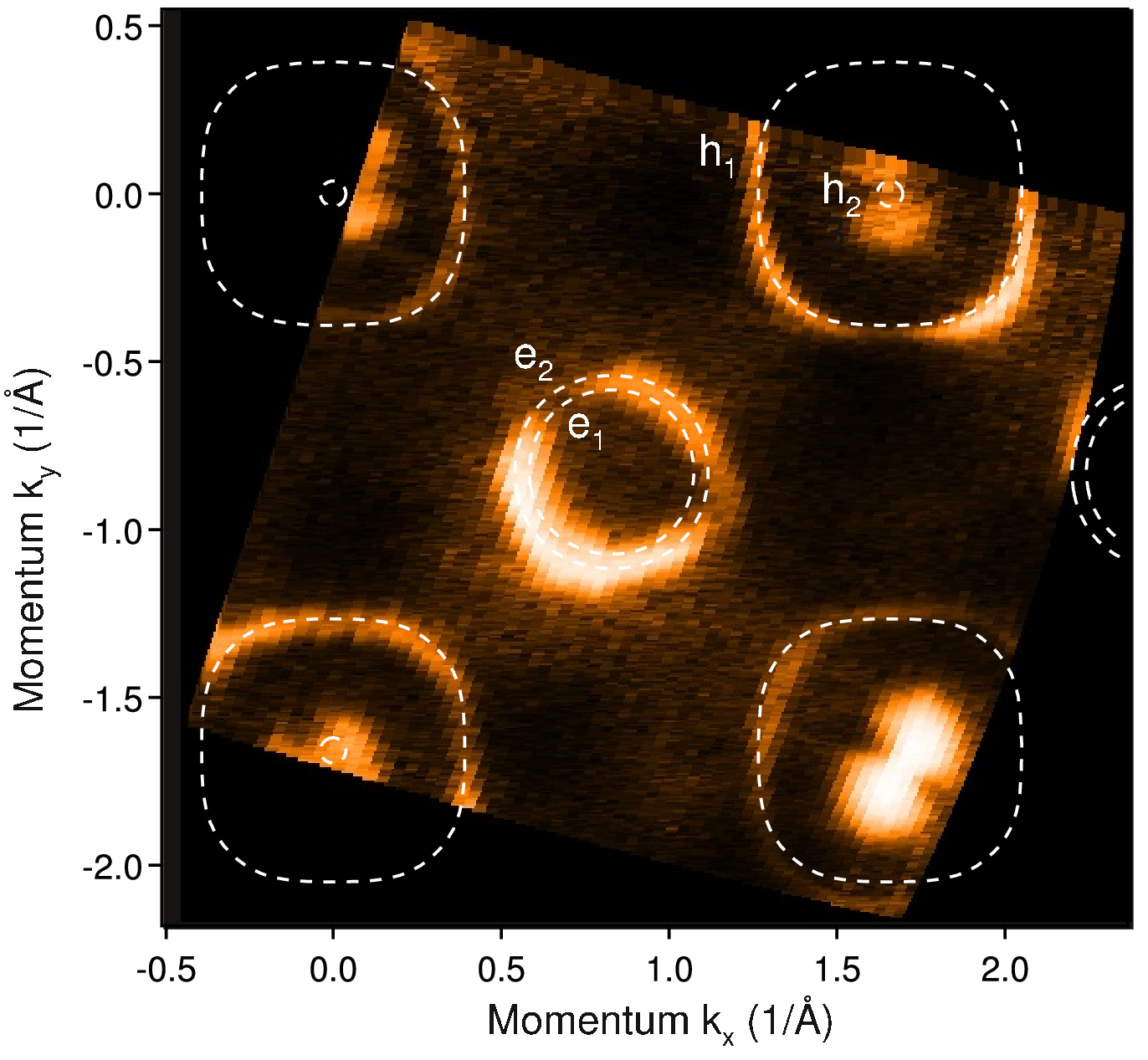}
  \caption{(color online) FS of a LiFeAs from angle resolved photoemission.}
  \label{fig:FSexp}
\end{figure}

%%\textcolor{red}{Paragraph on ARPES...?}
Recent ARPES experiments on LiFeAs have revealed several interesting
features. Although the low-energy electronic structure is qualitatively
similar to other Fe-pnictides, % being formed by the five bands
the Fermi surface of LiFeAs is remarkably different. It shows a
strongly reduced tendency to ($\pi$, $\pi$)-nesting\cite{borisenko},
which is present in most of the new iron superconductors and often
believed to be characteristic for all of them. As shown in
Fig.~\ref{fig:FSexp}, the radii of the hole and electron cylinders are
quite different here.  Another observation is the renormalization of the
conduction bands by a factor up to 3 as compared to DFT calculations
with an extended van Hove singularity (flat maximum of the h$_2$ band)
in the center of the Brillouin zone. The four bands supporting the Fermi
surface were found to be differently, though isotropically, gapped in
the superconducting state.\cite{borisenko} These observations can have
important implications for understanding the mechanism of
superconductivity in the Fe-pnictides. Here we show that there is no
surface driven electronic structure in LiFeAs and therefore surface
sensitive measurements represent the bulk electronic structure as
opposed to the situation in the 1111 (e.g. LaOFeAs) compounds.
%Therefore it is important to
%understand whether this information is peculiar to the bulk of the
%material or merely characterizes the surface as it was the case with
%1111 compounds.

Non-spin-polarized scalar relativistic DFT total energy and Kohn-Sham band structure %
calculations were performed with the full-potential local-orbital code %
version FPLO9.00 \cite{koepernik}. All considered structures were fully %
relaxed with the generalized-gradient-approximation (GGA) functional \cite{GGA} %
using the force tool of FPLO9.00. Forces were minimized to less then $5\cdot10^{-3}$~eV/\AA~%
for bulk and $10^{-1}$~eV/\AA~for slab calculations. %
This is necessary since no detailed surface structure is known from experiment. %

The calculations were carefully converged with the number of k-points, and %
finally 12,12,4 regular k-point meshes were used in the full Brillouin zone (BZ). %
The densities were converged down to $10^{-6}$ for the corresponding parameter %
in FPLO ($\text{rms}(\rho^\text{out}-\rho^\text{in})/\text{u.c.}$). Table~\ref{tab:1.1}~shows the results for the bulk lattice parameters %
compared to experiment \cite{tappl,chua,pitcher}.%

One can observe a decrease in volume of the cell, compared to
experimental data, which is reflected in a $c$-lattice
parameter reduction %
due to the softness of the lattice in $z$-direction (cf. LaOFeAs in
Ref.~\onlinecite{eschrig2}). This slight underestimation of the lattice
parameters even in GGA calculations is a common feature for
non-spin-polarized calculations in the pnictide family due to the strong
coupling of magnetism to the lattice.  However, %
the differences remain acceptable %
although the deviation for the distance in $z$-direction between adjacent
Fe and As layers d$_{\text{Fe-As}}$ is approximately 3\%. %
The local density approximation (LDA), would result in even further %
reduced $a$- and $c$-values (results not shown), %
hence GGA was used throughout this work.%

\begin{table}%[!h]
  \centering
  \begin{tabular}{|l|c|c|c|c|c|}
    \hline
    		& & & & & \\[-2.3ex]
            &$\quad a[$\AA$] \quad$&$\quad c[$\AA$] \quad$&
               $\quad z_{\text{As}} \quad$&$\quad z_{\text{Li}} \quad$
			&$\quad d_{\text{Fe-As}}[$\AA$] \quad$               \\[0.2ex]
	\hline
    experim.  &  3.791  &  6.364  &  0.2635 &  0.8459 & 2.420 \\
    GGA       &  3.766  &  6.158  &  0.2729 &  0.8325 & 2.345 \\\hline
  \end{tabular}
  \caption{Structure parameters of bulk LiFeAs; d$_{\text{Fe-As}}$ is the %
  atom layer distance in $z$-direction; experimental data taken from Ref.~\onlinecite{chua}.
  \label{tab:1.1}}
\end{table}

After relaxation, the next step was to calculate the cleaving %
behavior of LiFeAs by the same approach as in
Ref.~\onlinecite{eschrig2}. %
We had to use an orthorhombic Pmm2 space group instead %
of P4/nmm, in order to allow all atom layers to relax in $z$-direction, unconstrained by symmetry. 

In order to study the cleaving behavior we constructed a super-cell with
two unit cells in $z$-direction, with lattice constant $2c$. For
$c=c_\text{bulk}$ this just represents a bulk calculation. Now, we apply
tensile strain in $z$-direction, by increasing $c$. We performed a set
of calculations for $c$-values, which correspond to 0, 10, 20 and 30$\%$
strain. 

\begin{figure}%[!ht]
 \centering
  \includegraphics[angle=0,scale=0.40,clip=]{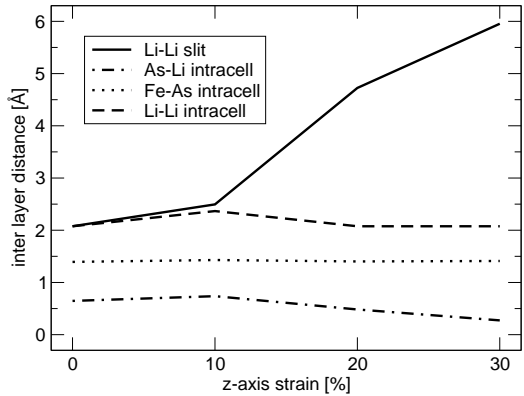}
  \caption{Inter layer distances of LiFeAs as a function of tensile 
    strain in $z$-direction. 
    The lines between data points are guides for the eye.}
  \label{fig:asym}
\end{figure}

Fig. \ref{fig:asym} shows the inter and intra unit-cell layer distances
of different atomic layers, such as Li-Li (straight and dashed lines), %
Fe-As (dotted line) and Li-As (dash-dot line) for this periodic
superstructure of two unit cells in $z$-direction as a function of
tensile strain. At large enough strain the original periodicity of
period $c$ in $z$-direction is broken by forming slits. By our adopted
superstructure of two unit cells, slits form between every other Li-Li
layer pair.  (\textit{Inter} indicates the %
distance crossing the slit, when it is formed, while \textit{intra} %
corresponds to distances within the two unit cell slab
$[Li|As|Fe_2|As|Li|Li|As|Fe_2|As|Li]$.) %

For small strain the whole slab is more or less uniformly stretched with
all the layer distances increasing compared to their bulk values.  From
the calculated forces of about 0.2~eV/\AA~one infers a tensile stress in
z-direction of 2.1~GPa for 10\% tensile strain.  For a strain between 10
and 20$\%$,  LiFeAs cleaves between two Li layers in our
calculations with a resulting neutral surface and consequently a
non-polar slab, as we pointed out above (see also
Ref.~\onlinecite{borisenko}).  Cleavage here means that the layers are
no longer uniformly stretched but that the two unit cells relax back to
bulk like distances, while a slit (vacuum) opens up between two Li
layers such that we have a repeated slab setup with a slab thickness of
two unit cells.  In the doubled unit cell the interior Li-Li layer
distance (termed \textit{intracell} in Fig.~\ref{fig:asym}) relaxes back
even below the bulk value. Also the topmost Li-As layer distance close
to the slit reduces appreciably while the Fe-As layer distance hardly
changes. (It should be pointed out that our simulation of the cleaving process only 
provides its qualitative behavior, because the chosen supercell
may not be large enough to simulate true bulk like behavior in the
interior layers.)

\begin{table}%[ht]
\begin{tabular}{|c|c|c||c|c|}
\hline 
		& & & & \\[-2.3ex]
 		  & bulk (4uc) & 3/1 slab & 3/3 slab & bulk (6uc) \\
\hline
	$\ldots$ & $\ldots$ & Vac. & Vac. & $\ldots$ \\	
	Li/As &	0.650 & 0.496 & 0.498 & 0.650\\
	As/Fe &	1.398 & 1.425 & 1.424 & 1.398\\
	Fe/As &	1.398 & 1.396 & 1.396 & 1.398\\
	As/Li &	0.650 & 0.634 & 0.636 & 0.650\\
	Li/Li &	2.062 & 2.065 & 2.066 & 2.062\\
	Li/As &	0.650 & 0.643 & 0.644 & 0.650\\
	As/Fe &	1.398 & 1.398 & 1.399 & 1.398\\
	$\cdots$ & $\cdots$ &  $\cdots$ & $\cdots$ & $\cdots$ \\
\hline
\hline 
%		& & & &  \\[-2.3ex]
 %	layer & bulk (4uc) & 3/1 slab & 3/3 slab & bulk (6uc) \\
%\hline
	$\ldots$ & $\ldots$	&  Vac. & Vac.   & $\ldots$ \\	
	Li 		& $-$0.144	& $-$0.161	& $-$0.162& $-$0.144\\
	As		& +0.296	& +0.426	& +0.426	& +0.296\\
	$Fe_2$	& $-$0.304	& $-$0.375	& $-$0.380	& $-$0.304\\
	As 		& +0.296	& +0.269	& +0.267	& +0.296\\
	Li		& $-$0.144	& $-$0.155	& $-$0.152& $-$0.144\\
	Li 		& $-$0.144	& $-$0.146	& $-$0.144	& $-$0.144\\
	As		& +0.296	& +0.295	& +0.293	& +0.296 \\
	$Fe_2$	& $-$0.304	&  $-$0.304	& $-$0.297	& $-$0.304\\
	$\cdots$ &  $\cdots$	& $\cdots$	& $\cdots$  & $\cdots$	  \\
\hline
\end{tabular}
\caption{Upper part: Inter layer distances in \AA . Lower part: 
  Electron excess per site. 
  Left and right double column: see explanation in the
  text. In view of the mirror symmetry with respect to the central layer only
  half of a slab is represented.}
\label{tab:dist}
\end{table}

To obtain a reliable lattice and electronic structure of a cleaved crystal %
surface it is necessary to choose a slab %
that is thick enough to pin the Fermi level of the slab to the bulk value %
and to choose a slit that is wide enough so that electronic wave functions %
from both sides of the slit do not overlap. (Due to the chosen mirror symmetry %
in  the center of the slab (z-direction) there is no electric field across the slit.) %
It turns out that the Fermi level of even a slab thickness of only three %
unit cells hardly changes compared to the bulk, indicating that this %
thickness is already sufficient. %
To check convergence with the vacuum layer width, calculations were done with a %
slit width of about $c_\text{bulk}$ (resulting in a periodicity 4c for the repeated three unit cells slab; termed 3/1 in Table~\ref{tab:dist}) and with a slit width of %
about $3c$ (resulting in a periodicity of 6c for the whole repeated slab setup, termed 3/3). %
In order to have $\vec{k}$-integration errors under control, the bulk %
calculations (used for comparison in Table~\ref{tab:dist}) were done with supercells of the same periodicity %
in $z$-direction (4 and 6 unit cells, respectively) allowing for the %
same $\vec{k}$-mesh interpolation in both bulk and slab cases. % 
The two different vacuum setups with their corresponding bulk
calculations
are represented by the left and right double columns in Table ~\ref{tab:dist}.

When comparing the inner layer distances and layer charges %
of the slab with those of the bulk crystal (lower parts of both sub
panels in %
Table~\ref{tab:dist}), one sees that a slab of three unit cells %
thickness is already giving bulk representative interior layers. %
From the residual forces the numerical accuracy of the distances is
estimated to be better than 0.01\AA, while the accuracy of the layer charges
is better than 0.005 of the electron charge. 
Also the positions and charges between a calculation with a vacuum
width of $c$ and $3c$ barely differ, so that in what follows the 3/1 slab %
is used for the surface band structure analysis.%

As is seen from the upper parts of Table~\ref{tab:dist}, there is a slight %
outward shift of the topmost As layer beneath the surface and an appreciable %
inward shift of the Li surface layer. %
Interestingly, this causes a charge redistribution between the As layer %
and the Fe layer below of about 0.1 electron charge, but nearly no change %
of the Li layer charge.%

\begin{figure}%[!ht]
  \centering
  \includegraphics[angle=-90,scale=0.31,clip=]{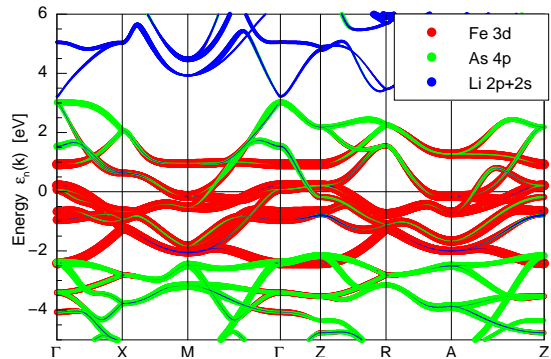}
  \caption{Band structure of bulk LiFeAs, orbital projected
    onto the leading Fe, As and Li orbitals.}
  \label{fig:bandplot}
\end{figure}

\begin{figure}%[ht]
  \centering
  \includegraphics[angle=-90,scale=0.31,clip=]{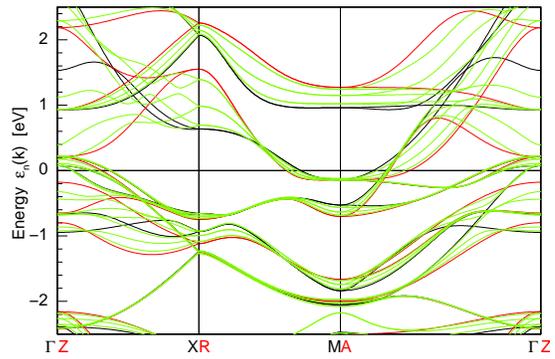}
  \caption{Band plot of bulk LiFeAs for $k_z$=~0 (black) %
  			and $k_z$=$~\pi$ (red) and slab LiFeAs (green lines).}
  \label{fig:bulk}
\end{figure}

\begin{figure}%[ht]
  \centering
  \includegraphics[scale=0.2,clip=]{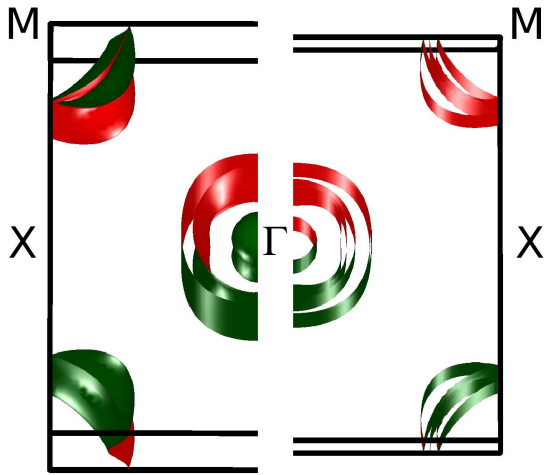}
  \caption{(color online) FS of LiFeAs. The Z-point is above $\Gamma$,
  			the A-point above M and, the R-point above X.
Left panel: bulk. Right panel: slab (no $z$-dispersion; 3D perspective
for better visibility only).}
  \label{fig:FSbulk}
\end{figure}

After having justified our model, we will inspect more closely
Figs.~\ref{fig:bandplot} to \ref{fig:FSbulk}, which show results for the
band-structure and Fermi surface (FS).  In Fig.~\ref{fig:bandplot} we
present the orbital projected band structure for the bulk calculation.
The orbital projection is obtained by projecting the Kohn-Sham wave
functions onto the chemical orbitals of the FPLO basis.  The line width of
the colored lines in Fig.~\ref{fig:bandplot} indicates the orbital
content/character in the wave function for every band.  As is well known
the Fe 3d bands (shown in red) together with some hybridization of the
As 4p states (green) dominate the entire low energy spectrum. This is
responsible for the metallic behavior of LiFeAs.

However, it is important to note the strong $z$-dispersion of a Li band
(shown in blue), coming from about 5~eV between M and $\Gamma$ down to
$\Gamma$ (at 3eV) and going further down to \mbox{-2~eV} via several
branches along $\Gamma$-M and $\Gamma$-Z-R-A, which
couples to As at higher energies and to Fe at lower energies.  This
effect is absent in all other families of iron pnictides or
chalcogenides i.e., 11 (e.g. FeSe), 1111 (e.g. LaOFeAs) or 122 (e.g.
BaFe$_2$As$_2$)\cite{eschrig1}.  In fact, this dispersive feature is
responsible for the closed Fermi surface (FS) seen around $\Gamma$ in
Fig.~\ref{fig:FSbulk} (left panel). Of course, in a slab model there is
no $k_z$-dispersion and hence this FS is not closed in
Fig.~\ref{fig:FSbulk} (right panel).

As seen in Fig.~\ref{fig:bulk}, all slab bands (thin green lines) in the
vicinity of the Fermi level are within the $k_z$-dispersion range of the
corresponding bulk bands (pairs of black and red lines). Hence no
surface band is found in an energy window free of bulk bands.  This is
in contrast to the situation found for LaOFeAs in
Ref~\onlinecite{eschrig2}.  Furthermore, the bulk FS nesting is less
pronounced than in LaOFeAs (larger difference between radii of electron
and hole FSs), due to a larger $k_z$-dispersion compared to LaOFeAs,
which is mainly caused by the extended Li states.  This could be related
to the absence of magnetic order in this material.
 
We note that Ref.~\onlinecite{singh} calculated a smaller
orbital weight ratio between As 4p and Fe 3d. Although both calculations
differ somewhat in the Fe-As layer distance (here 2.345~\AA~compared to
2.403~\AA~of Ref.~\onlinecite{singh}) this is mainly caused by different
local orbital projections.

Fig.~\ref{fig:FSbulk} shows the calculated FSs of the bulk and the slab.
As can be seen the $k_z$-dependence of the bulk FSs interpolates between
the corresponding $k_z$-dispersionless slab FSs, and no indication of
surface specific features is found. Hence, surface sensitive experiments
will represent the bulk electronic structure.

ARPES results for the FS are shown on Fig.~\ref{fig:FSexp}. The FS radii are
comparable with the calculated results at least in chosen directions
while the shape of the FSs around $\Gamma$ seems to be more complex in
experiment. We infer that this finding, just as the band renormalization
presented below, is intrinsic to the bulk electronic structure.

In Table~\ref{tab:FSdata} the calculated bulk %
Fermi radii (averaged over $k_z$) are compared to measured values (see Fig.~\ref{fig:FSexp}).

\begin{table}%[!h]
  \centering
  \begin{tabular}{|l|c|c|c|c|}
    \hline
    		& & & & \\[-2.3ex]
            &$\quad M_1 \quad$&$\quad M_2 \quad$&
%               $\quad \mid\hspace{-3pt}M\hspace{-3pt}\mid \quad$&$\quad
%               \Gamma_1 \quad$
               $\quad \Gamma_1 \quad$
			&$\quad \Gamma_2 \quad$               \\
	\hline
    experim.  &  0.15  &  0.18    & 0.24 &  $\approx 0.07$\\
    GGA       &  0.17  &  0.17    & 0.19 &  0.13\\\hline
  \end{tabular}
  \caption{Fermi radii in $k_x$-direction of different FS-pockets in
    units of $\frac{2\pi}{a}$ from experimental data and from the calculation.
}
  \label{tab:FSdata}
\end{table}

\begin{figure}%[!ht]
  \begin{center}
  \includegraphics[angle=0,scale=0.9,clip=]{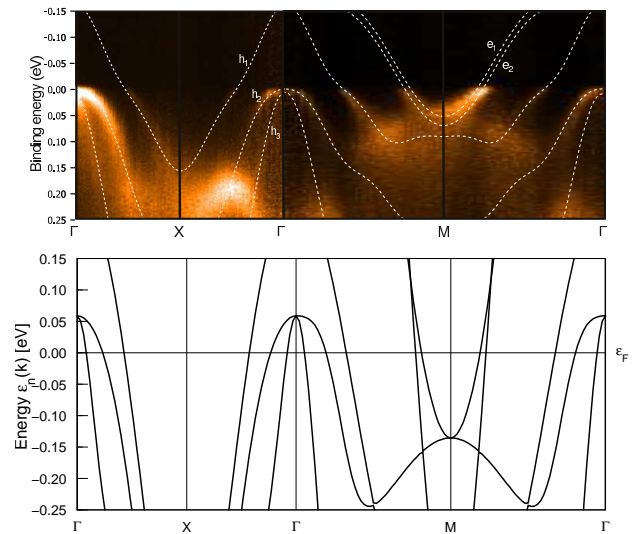}
  \end{center}
  \caption{ (color online) Upper panel. Photoemission intensity recorded
    along the high symmetry directions in LiFeAs. Lower panel.
    Corresponding results for the calculated bulk bandstructure.}
  \label{fig:exp_comp_theo}
\end{figure}

In Fig.~\ref{fig:exp_comp_theo} we compare the
experimental ARPES data with the results of the calculations. The spectra
were collected at BESSY. A detailed description of the experiment can be
found elsewhere (Ref.~\onlinecite{borisenko}).  In the upper panel, we
show the photoemission intensity and in the lower panel the
corresponding results for the calculated bulk band structure.  The
photoemission data have been interpolated by a tight-binding (TB) fit
(denoted by the dotted lines). Note, however, that there are
uncertainties in TB fits, especially concerning band connectivities.
The hole-like structures near the $\Gamma$-point are indicated by h$_1$,
h$_2$ and h$_3$, and electron-like structures near the corner of the BZ
by e$_1$ and e$_2$. These notations are also used in
Fig.~\ref{fig:FSexp} (except for h$_3$, see below).

The energy scale in both panels is of the same width, which allows to
estimate the band renormalization.  The overall agreement is remarkable,
taking into account that in both cases the low energy electronic
structure is formed by five bands and the Fermi surface (FS) is of the
same topology. Applying a renormalization factor to the calculations
leads to even better, quantitative agreement. All characteristic
dispersions and Fermi velocities are reproduced by the calculations. The
experimental Fermi surface (Fig.~\ref{fig:FSexp}) consists of two
hole-like FSs around the $\Gamma$-point and two electron-like FSs
centered at the M-point. The strongly dispersive h$_3$ feature (seen in
Fig.~\ref{fig:exp_comp_theo}) barely contributes to the spectral weight
at the Fermi level and hence is not indicated in Fig.~\ref{fig:FSexp}.
The relative sizes of the Fermi surfaces differ slightly from the
theoretical ones. Notably, the first two hole pockets at the zone
center seem to be smaller in the ARPES data compared to the
calculations. (The agreement can be improved by applying a small shift
of the chemical potential. This would also improve the agreement
concerning the h$_3$ feature at the $\Gamma$-point.)

In summary, we have investigated the surface structure and surface effects on the
electronic structure in LiFeAs.  The main finding is that
due to the neutral cleaving of the compound between the Li-layers there
are no significant surface effects on the electronic states. Consequently,
LiFeAs is an ideal compound in the pnictide family for the study by
ARPES and STM methods.

\begin{acknowledgements} 
The project was supported, in part, by the DFG under Grants No. KN393/4, BO
1912/2-1 as well as priority program SPP1458.
\end{acknowledgements}

%\clearpage

\end{document}